
\magnification=1200
\parindent 1.0truecm
\baselineskip=16pt
\hsize=6.0truein
\vsize=8.5truein
\rm
\null


\footline={\hfil}
\vglue 0.8truecm
\rightline{\tt nucl-th/yymmxxx}
\rightline{\bf DFUPG-06-PG-01}
\rightline{\sl (last revised May 2007)}
\vglue 2.0truecm

\centerline{\bf  IS THE PENTAQUARK THE ONLY JUSTIFICATION }
\centerline{\bf FOR RESEARCH ON {\sl KN} PHYSICS ? }
\vglue 0.5truecm
\centerline{ Presented at {\sl "EURIDICE: The Final Meeting.}}
\centerline{\sl Effective Theories of Colours and Flavours: from 
EURODA$\Phi$NE to EURIDICE",}
\centerline{ Kazimierz, Poland, August 24 -- 27, 2006 }
\vglue 1.0truecm

\centerline{\sl P.M. Gensini$^{\dag}$, R. Hurtado$^{\star\star}$, Y.N. 
Srivastava$^{\dag}$ and G. Violini$^{\star\ddag}$ }
\vglue 0.3truecm
\centerline{\sl $^{\dag}$) Dip. di Fisica dell'Universit\`a di Perugia, Italy,}
\centerline{\sl and I.N.F.N., Sezione di Perugia, Italy.}
\centerline{\sl $^{\star}$) Dip. di Fisica dell'Universit\`a della Calabria, 
Arcavacata di Rende (Cosenza), Italy,}
\centerline{\sl and I.N.F.N., Laboratori Nazionali di Frascati, Gruppo 
Collegato di Cosenza, Italy.}
\centerline{\sl $^{\star\star}$) Departamento de F{\'\i}sica, Universidad 
Nacional de Colombia, Ciudad Universitaria, }
\centerline{\sl Carrera 30 No. 45--03, Edificio 405, Oficina 218, Bogot\'a, 
Colombia .}
\centerline{ $^{\ddag}$) Speaker at the meeting.}
\vglue 1.0truecm

\centerline{\bf ABSTRACT}

The talk is intended to motivate the use of DA$\Phi$NE--2 running at the 
$\phi$ peak as an intense, clean source of low--momentum charged and neutral 
kaons. It covers a few open problems still unsolved after more than 
twenty--five years and the physics (some of it still novel) that could be 
learned only in this way. And, of course, the answer to the question in the 
title is {\sl NO}.

\pageno=0
\vfill
\eject


\footline={\hss\tenrm\folio\hss}

\vglue 1.0truecm
\centerline{\bf Is the Pentaquark the Only Justification for Research on {\sl 
KN} Physics ? }
\vglue 0.5truecm
\centerline{\sl P.M. Gensini, R. Hurtado, Y.N. Srivastava and 
G. Violini }

\vglue 1.0truecm
\noindent{\bf 1. Introduction: about history (and philosophy).}
\vglue 0.3truecm

In the last few years the interest for kaon physics has significantly 
increased. The possibly most spectacular reason for this revival of 
interest for the understanding of kaon--nucleon interaction has been the 
suggestion of the possible existence of a pentaquark reported in 2003 by 
the LEPS/SPring8 group in Japan [1]. However, after a couple of 
years new experimental results [2] cast serious doubts [3] on the 
existence of such a state.

The fact that usually one makes reference to a theoretical prediction about 
the possible existence of such a state [4] might cover the fact that 
experimentally there had already been some indication of that sort, since 
about thirty years ago a bump in $K^+N$ total cross sections in the 1 GeV/c 
region prompted much interest about the possibility of existence of a resonant 
state which would not fit a classification in either an {\bf 8} or {\bf 10} 
representation of SU(3)$_f$. Investigations by phase--shift analyses of $K^+N$ 
scattering did not lead to a conclusive evidence, although some of the 
solutions actually exhibited a resonant behaviour in the energy 
region where the ``pentaquark'' was supposed to lie [5]. It has 
however to be remarked that the widths of those putative resonances were 
significantly larger than the one usually assigned to the pentaquark, although 
a recent analysis of $K^+N$ has suggested that only a very small width would 
be compatible with the data [6].

It is to be noted that, having those old results been almost forgotten, the 
claim of the discovery of a pentaquark was something unexpected, 
and although linked to a theoretical prediction, it was not the result of a 
dedicated, systematic search. One could argue that if in the eighties the 
machines producing medium-energy (for the scales of that time) kaons 
had not been turned off, probably the consequent deeper knowledge of $KN$ 
interaction could have helped in understanding better such an unexpected (at 
least by most) phenomenon. This is recognized in Hicks' review [7] 
and it is interesting to note that it also underlines the fact that for this 
purpose even the (relatively) good $K^+N$ data require new, better experiments.

A different situation is related to the increasing attention to kaon physics 
originated by the starting of operation at DA$\Phi$NE, where a systematic 
research program is carried out. 

As it is well known, during the first phase of DA$\Phi$NE's activity three 
experiments have been performed. On the ground of fundamental physics, KLOE 
seems to be the one proging more basic problems, since its goal is the study 
of tiny effects of CP violation in the decays of neutral kaons. The other two, 
FINUDA and DEAR, are devoted to the study of, respectively, hypernuclei 
and kaonic atoms; but, for what concerns FINUDA, one should recall its 
possibility of taking $K_L$ charge exchange data on the hydrogen of its 
plastic scintillators, following a proposal by Olin [8].

One of the main results of DEAR, namely the solution of the long--standing 
puzzle of the character of low--energy $K^-p$ interaction, giving a definite 
confirmation that it is repulsive [9], in agreement with all the 
analyses of the avialable low--energy $K^-N$ data, has a strict connection 
with the main aspect of this talk. 

The analogy between this and the pentaquark issue is that they give a common 
lesson: the experimental knowledge we have of $KN$ physics at lab. momenta 
below 1 GeV/c is poor and based on old data. 

\vglue 1.0truecm
\noindent{\bf 2. A look into possible futures at DA$\Phi$NE.}
\vglue 0.3truecm

Even comparing at a glance $KN$ and $\pi N$ total cross sections [10] 
is enough to confirm this statement, and this fact reflects in turn on the 
knowledge of the 
parameters of the $KN$ interaction (scattering lengths, coupling constants, 
sigma terms), much worse that that of the SU(3)$_f$--related $\pi N$ ones.

One could argue that, despite this difference in quality, nevertheless it has 
been possible to analyze kaon data in a coherent way, extracting the relevant 
information, and describing it in terms of a few parameters: however, this is 
only partially true, because, for example, the calculation of $KN$ sigma terms 
by dispersive methods [11] is affected by substantial 
uncertainties, and the coupling constants involving strange particles have 
much larger errors than those of the $S = 0$ sector, so that the success of 
the comparison of their values with SU(3)$_f$ predictions, usually claimed in 
particle physics textbooks is not so evident. Table I offers an 
order--of--magnitude estimate of the uncertainties for several couplings 
accessible though dispersive analyses.

\vglue 1.0truecm
\centerline{\bf Table I$^\star$}
\vglue 0.3truecm

\hrule

$$\vbox{\halign{#\hfil & \hfil#\hfil & \hfil#\hfil & \hfil#\hfil & \hfil#\hfil 
& \hfil#\hfil \cr
& & & & & \cr
Coupling constant & $g^2_{\pi{NN}}$ & $g^2_{K\Lambda{p}}$ & $g^2_{K\Sigma{N}}$ 
& $g^2_{\pi\Lambda\Sigma}$ & $g^2_{\pi\Sigma\Sigma}$ \cr
& & & & & \cr
SU(3) prediction & $g^2$ & ${1\over3}(1+2\alpha)^2g^2$ & $(1-2\alpha)^2g^2$ & 
${4\over3}(1-\alpha)^2g^2$ & $4\alpha^2g^2$ \cr
uncertainty & a few \% & 10 \% & 30 \% & 100 \% & 100 \% \cr}}$$
\noindent $^\star$) Here $\alpha =f/(f+d)$ is a typical 
parameter of the theory, due to the existence of two {\bf 8} representations 
in the {\bf 8} $\otimes$ {\bf 8} product.
\hrule
\vglue 1.0truecm

The scope of this talk is to review the description of $KN$ interactions at 
low energies, and to put in evidence a number of problems which still exist 
and which can only be solved by new experiments, most of which are within the 
reach of DA$\Phi$NE.

We shall not give many technical details, since there are several papers by 
our group where they are exhaustively presented [12]. Our purpose is to 
show, mainly to our experimental colleagues, that with a little effort one 
could have a much better understanding of this branch of physics. It is 
interesting to note that indeed some experimental proposals for the future of 
DA$\Phi$NE are taking into account these ideas [13,14].

$KN$ reactions are described by four isospin amplitudes, two for each 
strange\-ness sector. The $S = +1$ sector is well described by an $S$--wave 
scattering length approximation in both isospin channels (see Table 2), being 
the $P$--wave significant only in the $I = 0$ channel from about 300 MeV/c on.

\vglue 1.0truecm
\centerline{\bf Table II }
\vglue 0.3truecm

\hrule
$$\vbox{\halign{#\hfil & \hfil#\hfil \cr
& \cr
I = 1 & about -0.3 {\sl fm} (minor variations if an effective range is 
included) \cr
I = 0 & very small (between -0.1 and 0.2 {\sl fm}) \cr}}$$

\hrule
\vglue 1.0truecm

The situation is much more complicated for the $S = -1$ sector, due to 
the presence of several coupled channels. Some fifty years ago Dalitz and Tuan 
proposed a formalism that in its simplest application (scattering lengths) 
succeeded in predicting the existence of a resonance below the elastic 
threshold, the $\Lambda$(1405) [15]. Few years later, a more 
complicated multichannel version of this formalism including $S$--, $P$-- and 
$D$--waves was used to analyze data up to about .5 GeV/c [16]. As of 
today, this latter is one of the best, model--free analyses available for 
these systems$^{\dag}$.

A characteristic of this formalism is that the continuation of the 
parame\-trization to the unphysical regions automatically includes the correct 
theoretical behaviour at $\pi\Lambda$ and $\pi\Sigma$ thresholds.

The understanding of the interaction in the low--energy region is not exempt 
of problems, and this cannot be surprising insofar it is evident that no 
formalism can replace the scarce quality of (or even the lack of) the 
experimental data it aims to describe.

Before going to mention some of these problems, it is appropriate to recall 
that the lowest energy where (poor) data exist lies tens of MeV/c above the 
region that could be studied using DA$\Phi$NE kaons.

The first problem we would like to mention, put in evidence a few years ago 
by some of us, is that dispersion relations for $\pi Y$ scattering indicate 
that something might go wrong in Kim's multichannel parametrization.

For the youngest colleagues who may be not too familiar with this tool broadly 
used in $KN$ physics during the sixties and seventies, we recall that the 
analyticity of the scattering amplitude as function of the energy can be used 
not only to test the consistency of the values of the forward differential 
cross sections with the total cross sections (and through this the validity of 
causality at short distances), but also to determine the values of the coupling 
constants of the particles involved [17]. This application has a long 
history in the case of $KN$ physics, where it was used as a test of 
SU(3)$_f$ symmetry, and, as we have shown in Table I, the different quality of 
pion and kaon data shows up in the relative uncertainties of the corresponding 
couplings.

Going back to $\pi Y$ interactions, one can use the $\pi Y$ amplitudes 
provided by the multichannel parametrization of $S = -1$ $KN$ scattering to 
determine, by conventional dispersion relations, the values of the $\pi Y Y'$ 
couplings [18]. Their values, far from being constant, turned out to 
depend quite strongly on the energy at which the relations were evaluated: 
this behaviour was clearly signaling that something was not all--right either 
with the method (which however was quite successfull in all other cases) or 
with at least one of the higher--$\ell$ partial waves. Figures 1 through 3 
summarise nicely those results.

A second problem concerns the characteristic feature of $\bar{K}N$ system, 
namely the existence of $S = -1$ resonances below the elastic threshold, the 
$\Sigma$(1385) and the $\Lambda$(1405). Our knowledge about them is limited, 
and comes mostly from production experiments and only in part from the 
extrapolation below threshold of the low--energy  $\bar{K}N$ data. It must be 
observed that this region is inaccessible only to scattering experiments on 
hydrogen, but can be explored either in associate production or 
by experiments on nuclear targets, when part of 
the incoming kaon momentum can be carried out by the spectator nucleons 
[19]. For $^4$He (the gas filling KLOE's wire chamber), final state 
interactions {\sl in the inelastic channels} should not be a taxing problem 
due to the weak binding in nuclear states with $A \leq$ 3.

Because of the possibility of exploring deeply the unphysical regions, 
experiments on nuclei would allow to improve our knowledge of the 
$\Sigma$(1385) and $\Lambda$(1405) resonances, and particularly to clarify the 
nature of the latter, on which much discussion exists in the literature, and 
it has even been proposed the possibility that it is actually the result of 
the confluence of two resonant states [20]; it is to be remarked 
however that the only {\sl phenomenological} support to this hypothesis 
comes from a poor analysis [21] of a low--statistics 
experiment [22], and that related measurements could be performed 
with much higher statistics at DA$\Phi$NE.

Recently, two groups [23.24] have investigated this matter and 
the consistency of $K^- p$ scattering length with KEK [25] and DEAR 
[9] measurements of the $1s$ $K^- p$ atomic level shift. We shall not 
insist again on the fact that the experiments that led to attribute attractive 
character to the $K^- N$ low-energy interactions go again back to the 
infamous eighties [26].

Both these groups make use of an approach based on chiral SU(3) symmetry, 
and their results leave still open several puzzling questions. In particular, 
Oller {\sl et al.} [23] find two classes of solutions, one of which 
disagrees with DEAR measurements [9] (even if it is compatible with 
the less accurate KEK data [25]). Borasoy {\sl et al.} [24] 
criticize the consistency of the first solution with fundamental principles 
of scattering theory and prefer a KEK type solution.

These studies do not therefore question the repulsive character of the 
interaction, yet they suggest a reflection. The idea of using theoretical 
constraints in $K^- N$ analyses is not new, and it can be found in the 
literature in many variants (see, for instance, ref. [27]); in 
particular it is implicit in a list of several current elements of interest 
for these reactions, among which chiral symmetry is quoted in the first place 
[28].

One can always try to constrain a fit by imposing the validity of the 
hypotheses to be eventually tested: however, in this way one is 
substituting the knowledge of experimental data of adequate quality and 
statistics with a theoretical (possibly well founded, but still 
theoretical) prejudice. This is not accidental, because this branch of 
physics has been plagued by the absence of new experimental results for more 
than twenty years, during which theoretical research has made much progress, 
especially with low--energy, QCD--inspired methods. Our point of view is that 
the desirable, sounder procedure would be to try and gain better experimental 
data, that could be used to test the validity of any given approach.

As a matter of fact, it is clear that new good experiments can easily 
provide better and more abundant data than those which, {\sl faute de mieux},  
could be used for example by Oller {\sl et al.} (94 data points, referring to 
six reactions and in a very limited energy region). Obviously this does 
not pretend to be a criticism to Oller's approach, but only a reminder that 
the scarcity of data on $KN$ scattering is a direct consequence of the 
closing down of the machines where those data could have been produced.

In the last two decades a number of proposals of new facilities were debated 
(e.g. the European Hadron Facility [29] and KAON at TRIUMF [30]), 
but did not materialize for several -- even political -- reasons, and the few 
remaining kaon beams were barely enough to keep alive hypernuclear and 
exotic--atom physics.

One could still take advantage of the fact that, with the starting of 
DA$\Phi$NE's operations, the situation has changed, at least potentially, for 
the better. Although understandably the goals of the experiments 
carried out at DA$\Phi$NE during the first phase of its existence were not 
the improvement of our understanding of $KN$ interactions, our group has 
repeatedly [12,19] stressed that the experiments running there 
could also indirectly collect many events which could shed 
light on the above problems, from $K^-$ interactions and $K_L$ 
charge--exchange (and regeneration) both on $^4$He and H. Furthermore, 
DA$\Phi$NE (running at the $\phi$--resonance peak) is unique for exploring 
directly an energy region where otherwise the currently existing data would 
only allow to infer the behaviour of the scattering amplitudes via 
extrapolations from the higher--energy region. 

Indeed DA$\Phi$NE's  monochromatic charged (neutral) kaons are produced at 
momenta of about 127 (110) MeV/c, making possible (via the energy losses in 
the detector) to explore the region down to about 90 MeV/c, and there are 
at least two reasons for doing so. First, that region is sensitive to the 
details of the opening of the $\bar K^0 n$ channel; second, the possibility 
of collecting {\sl in the same experimental conditions} data from $K^+, K^-$ 
and $K^0_L$ allows for an accurate, simultaneous isotopic spin analysis of 
different reactions in either $S$ sector.

In fact, since $K^-N$ and $K^+N$ are described by four isospin amplitudes, the 
consideration of the charge exchange and regeneration amplitudes beside the 
elastic scattering amplitudes (which in principle are sufficient to determine 
completely the four amplitudes) leads to a set of overdetermined data 
(Table III).

A byproduct of this overdetermination is that the possibility of studying the 
regeneration on hydrogen would provide an information for a combination of 
$Kn$ amplitudes free of the need of taking into account the neutron Fermi 
motion [31]. Better regeneration data would also be able to improve 
considerably the determination of $g^2_{K\Sigma{N}}$ [32].

\vglue 1.0truecm
\centerline{\bf Table III }
\vglue 0.3truecm
\hrule

$$\vbox{\halign{\hfil#\hfil & \hfil#\hfil & \hfil#\hfil \cr
& & \cr
S = -1 & I = 0 & $K^- p - {1\over2} K^- n$ \cr
S = -1 & I = 1 & $K^- n$ \cr
S = +1 & I = 0 & $K^+ n - {1\over2} K^+ p$ \cr
S = +1 & I = 1 & $K^+ p$ \cr 
S = -1 & Ch.Exch. & $K^- p - K^- n$ \cr
S = +1 & Ch.Exch. & $K^+ p - K^+ n$ \cr
Regeneration & on H & $K^- n - K^+ n$ \cr}}$$

\hrule
\vglue 1.0truecm

In this connection one should observe that the interest for the region very 
close to elastic threshold may lead to overstating the importance of 
$S$--waves, but $P$--waves should not be neglected, and for several good 
reasons, such as the possibility of understanding of the 
nature of $\Lambda$(1405) through their interference with the $S$--waves, 
and of studying kaonic helium, an expected development of DEAR's program 
[33].

Until now we have insisted on the very low-energy region; however it should 
not be ignored that also the intermediate region is far from being fully 
understood. Beside the problem of the pentaquark, in that region one faces 
the problem of the many missing $\Sigma$ and $\Lambda$ states, and moreover 
the continuation of the most popular phase--shift analysis [34] is 
unable to reproduce the structure below threshold, so that its matching with 
the low--energy parametrizations is not exempt of ambiguities. 

Last but not least, one can expect new data from JPARC, as well as at very 
high energy from new accelerators [35], with secondary beams having 
energies of a few GeV, when these facilities will be operating.

In order to be prepared to reach a coherent description of at least the low 
energy interaction one should exploit DA$\Phi$NE: this will allow to reliably 
use such description to test theoretical models that possibly can be later 
incorporated in the fits (with the caveat that a clear distinction between 
experimental data and theoretical inputs should not be forgotten).

At DA$\Phi$NE one could expect about $10^7$ two--body and $10^5$ three--body 
final--state events per year. Even taking into account some reduction in 
these figures due to different causes of particle losses, the rates achieved 
would be orders of magnitude above those of the lowest--energy available data 
of thirty years ago, or of the few, more recent experiments. 
Moreover the possibility of studying by nuclear targets 
the region below threshold might allow the analyses to take effectively into 
account the existence of the $\pi\pi\Lambda$ channel, whose threshold is in 
that region.

The fact that the emphasis of this talk is on strong interactions should not 
prevent us from making an additional comment on the possibilites offered by 
DA$\Phi$NE in the area of radiative captures, where one can expect $10^4 - 
10^5$ events/year, and actually determine these B.R. for the $\Lambda$(1405).

In conclusion, we insist on the value of systematic research, that for 
kaon--nucleon physics would fill a serious gap of information: with 
DA$\Phi$NE at present or higher luminosity, operating at the $\phi$ peak, 
we would have a great opportunity to carry on a program of this kind and 
it would really be a -- perhaps unrecoverable -- loss if this opportunity 
were not fully exploited.

\vglue 1.0truecm
\noindent{\bf 4. Acknowledgements.}
\vglue 0.3truecm
One of us (G.V.) would like to express his gratitude to the Centro de 
Modelamiento Matem\'atico of the Universidad de Chile for its hospitality 
when the text of this talk was prepared.

\vfill
\eject

\vglue 1.0truecm
\centerline{\bf FIGURE CAPTIONS }
\vglue 0.3truecm

{\bf Figure 1} - $G_{\pi\Lambda\Sigma}^2/4\pi$ from $\pi\Lambda\to\pi\Lambda$

{\bf Figure 2} - $G_{\pi\Sigma\Sigma}^2/4\pi$ from $\pi\Sigma\to\pi\Sigma$ 
and $\pi\Lambda\to\pi\Lambda$

{\bf Figure 3} - $G_{\pi\Lambda\Sigma}G_{\pi\Sigma\Sigma}/4\pi$ from 
$\pi\Lambda\to\pi\Sigma$

\vglue 1.0truecm
\centerline{\bf REFERENCES and FOOTNOTES }
\vglue 0.3truecm

\item{$^{\dag}$} One of us (P.M.G.) happened to share exactly this point of 
view with none else but the late R.H. Dalitz at a breakfast in Frascati on the 
occasion of "DA$\Phi$NE '95".

\item{[1]} T. Nakano {\sl et al.}, {\sl Phys. Rev. Lett.} {\bf 91}, 012002 
(2003).

\item{[2]} M. Battaglieri {\sl et al.}, {\sl Phys. Rev. Lett.} {\bf 96}, 
042001 (2006).

\item{[3]} A.R. Dzierba, C.A. Meyer, A.P. Szczepaniak, {\sl J. Phys. Conf. 
Ser.} {\bf 9}, 192 (2005).

\item{[4]} D. Diakonov, V. Petrov, M. Polyakov, {\sl Z. Phys.} {\bf A 359}, 
305 (1997).

\item{[5]}A.T. Lea, B.R. Martin, G.D. Thompson, {\sl Nucl. Phys.} {\bf B 26}, 
413 (1971); B.C. Wilson {\sl et al.}, {\sl Nucl. Phys.} {\bf B 42}, 445 (1972).

\item{[6]} R.A. Arndt, I.I. Strakowski, R.L. Workman, {\sl Nucl. Phys.} {\bf A 
754}, 261 (2005).

\item{[7]} K. Hicks, {\sl Prog. Part. Nucl. Phys.} {\bf 55}, 647 (2005).

\item{[8]} A. Olin, {\sl "Workshop on Physics and Deterctors for DA$\Phi$NE 
'95"}, R. Baldini, F. Bossi, G. Capon, G. Pancheri eds., {\sl Frascati Phys. 
Ser.} {\bf 4}, 379 (1996).

\item{[9]} G. Beer {\sl et al.}, {\sl Phys. Rev. Lett.} {\bf 94}, 212302 
(2005).

\item{[10]} W.-M. Yao {\sl et al.} (Particle Data Group): {\sl J. Phys.} {\bf 
G 33}, 1 (2006).

\item{[11]} B. Di Claudio, A.M. Rodr{\'\i}guez Vargas, G. Violini: {\sl Z. 
Phys.} {\bf C 3}, 75 (1979); A.M. Rodr{\'\i}guez Vargas, G. Violini: {\sl Z. 
Phys.} {\bf C 4}, 135 (1980). For a review of this and other approaches, see: 
P.M. Gensini, {\sl $\pi N$ Newslett.} {\bf 6},) 21 (1992; {\sl "$KN$ Sigma 
Terms, Strangeness in the Nucleon, and DA$\Phi$NE"}, extended version of a 
talk presented at {\sl "1998 LNF Spring School"}, DFPUG-98-GEN-02, Perugia 
1998 (arXiv: hep-ex/9804344).

\item{[12]} P.M. Gensini, G. Violini, {\sl "Proc. of the Workshop on Science 
at the KAON Factory"}, D.R. Gill ed., TRIUMF, Vancouver 1991, Vol. 2, Sect. 
7.5; P.M. Gensini, G. Violini, {\sl "Perspectives on Theoretical Nuclear 
Physics"}, L. Bracci, {\sl et al.} eds., ETS, Pisa 1992, p. 162; P.M. 
Gensini, G. Violini, {\sl Rev. Col. F\'\i{s.}} {\bf 24}, 51 (1992) 
(arXiv: nucl-th/9210007); P.M. Gensini, {\sl "The Second DA$\Phi$NE Physics 
Handbook"}, L. Maiani, G. Pancheri, N. Paver eds., LNF, Frascati 1995, 
Vol. II, p. 739 (arXiv: hep-ph/9504024); P.M. Gensini, R. Hurtado, G. 
Violini, {\sl Genshikaku Kenky\=u} {\bf 48}, N. 4, 51 (1998) (arXiv: 
nucl-th/9804024).

\item{[13]} P. Buchler {\sl et al.} (Amadeus Collaboration), {\sl "Study of 
Deeply Bound Kaonic Nuclar States at DA$\Phi$NE--2"}, L.o.I. to I.N.F.N., 
Frascati, March 2006.

\item{[14]} F. Ambrosino {\sl et al.}, {\sl Eur. Phys. J.} {\bf C 50}, 729 
(2007) (arXiv: hep-ex/0603056).

\item{[15]} R.H. Dalitz, S.F. Tuan, {\sl Ann. Phys. (N.Y.)} {\bf 10}, 307 
(1960).

\item{[16]} J.K. Kim, {\sl Phys. Rev. Lett.} {\bf 14}, 29 (1965).

\item{[17]} N.M. Queen, G. Violini, {\sl "Dispersion Theory in High Energy 
Physics"}, McMillan, London 1974.

\item{[18]} P.M. Gensini, R. Hurtado, G. Violini, {\sl $\pi N$ Newslett.} {\bf 
13}, 291 (1997) (arXiv: nucl-th/9709023); P.M. Gensini, R. Hurtado, G. 
Violini, {\sl "Baryons '98"}, D.W. Menze, B.Ch. Metsch eds., World Scientific, 
Singapore 1999, p. 593 (arXiv: nucl-th/9811010).

\item{[19]} P.M. Gensini, G. Pancheri. N.Y. Srivastava, G. Violini, {\sl 
"Low--Energy Kaon--Nucleus Interactions at a $\phi$--Factory"}, 
DFUPG-05-PG-02, Perugia, March 2006 (arXiv: nucl-th/0603043).

\item{[20]} D. Jido, J.A. Oller, E. Oset, A. Ramos. U.G. Mei\ss{ner}, {\sl 
Nucl. Phys.} {\bf A 725}, 181 (2003); {\sl Nucl. Phys.} {\bf A 755}, 669 
(2005).

\item{[21]} V.K. Magas, E. Oset, A. Ramos, {\sl Phys. Rev. Lett.} {\bf 95}, 
052301 (2005).

\item{[22]} S. Prakhov {\sl et al.} (Crystal Ball Collab.), {\sl Phys. Rev.} 
{\bf C 70}, 034605 (2004).

\item{[23]} J.A. Oller, J. Prades, M. Verbeni, {\sl Phys. Rev. Lett.} {\bf 
95}, 172502 (2005); J.A. Oller, {\sl Eur. Phys. J.} {\bf A 28}, 63 (2006).

\item{[24]} B. Borasoy, R. Ni\ss{ler}, W. Weise, {\sl Phys. Rev. Lett.} {\bf 
96}, 192201 (2006); B. Borasoy, U.-G. Mei\ss{ner}, R. Ni\ss{ler}, {\sl Phys. 
Rev.} {\bf C 74}, 055201 (2006).

\item{[25]} M. Iwasaki {\sl et al.}, {\sl Phys. Rev. Lett.} {\bf 78}, 3067 
(1997).

\item{[26]} J.D. Davies {\sl et al.}, {\sl Phys. Lett.} {\bf 83 B}, 55 (1979); 
M. Izycki {\sl et al.}, {\sl Z. Phys.} {\bf A 297}, 11 (1980).

\item{[27]} R. Hurtado, {\sl Heavy Ion Phys.} {\bf 11}, 383 (2000).

\item{[28]} A. Olin, T.S. Park, {\sl Nucl. Phys} {\bf A 691}, 295 (2001).

\item{[29]} {\sl "Proposal for a European Hadron Facility"}, J.F. Crawford 
ed., report EHF 87-18, Trieste - Mainz, May 1987.

\item{[30]} See {\sl "Proc. of the Workshop on Science at the KAON Factory}, 
D.R. Gill ed., TRIUMF, Vancouver, B.C., 1981.

\item{[31]} M. Lusignoli, M. Restignoli, G. Violini, {\sl Phys. Lett.} {\bf 
24 B}, 295 (1967).

\item{[32]} G.K. Atkin, B. Di Claudio, G. Violini, N.M. Queen, {\sl Phys. 
Lett.} {\bf 95 B}, 447 (1980).

\item{[33]} C. Petrascu, {\sl The Kaonic Helium Case}, pres at {\sl "5th 
Italy--Japan Symposium: Recent Achievements and Perspectives in Nuclear 
Physics"}, 3 -- 7 November 2004, Naples, Italy.

\item{[34]} M.L. Gupta, R.A. Arndt, L.D. Roper, {\sl Nucl. Phys.} {\bf B 37}, 
173 (1972).

\item{[35]} M.G. Albrow {\sl et al.}, {\sl "Physics at a Fermilab Proton 
Driver"}, Fermilab Report, 2005 (arXiv: hep-ex/0509019).

\vfill
\eject

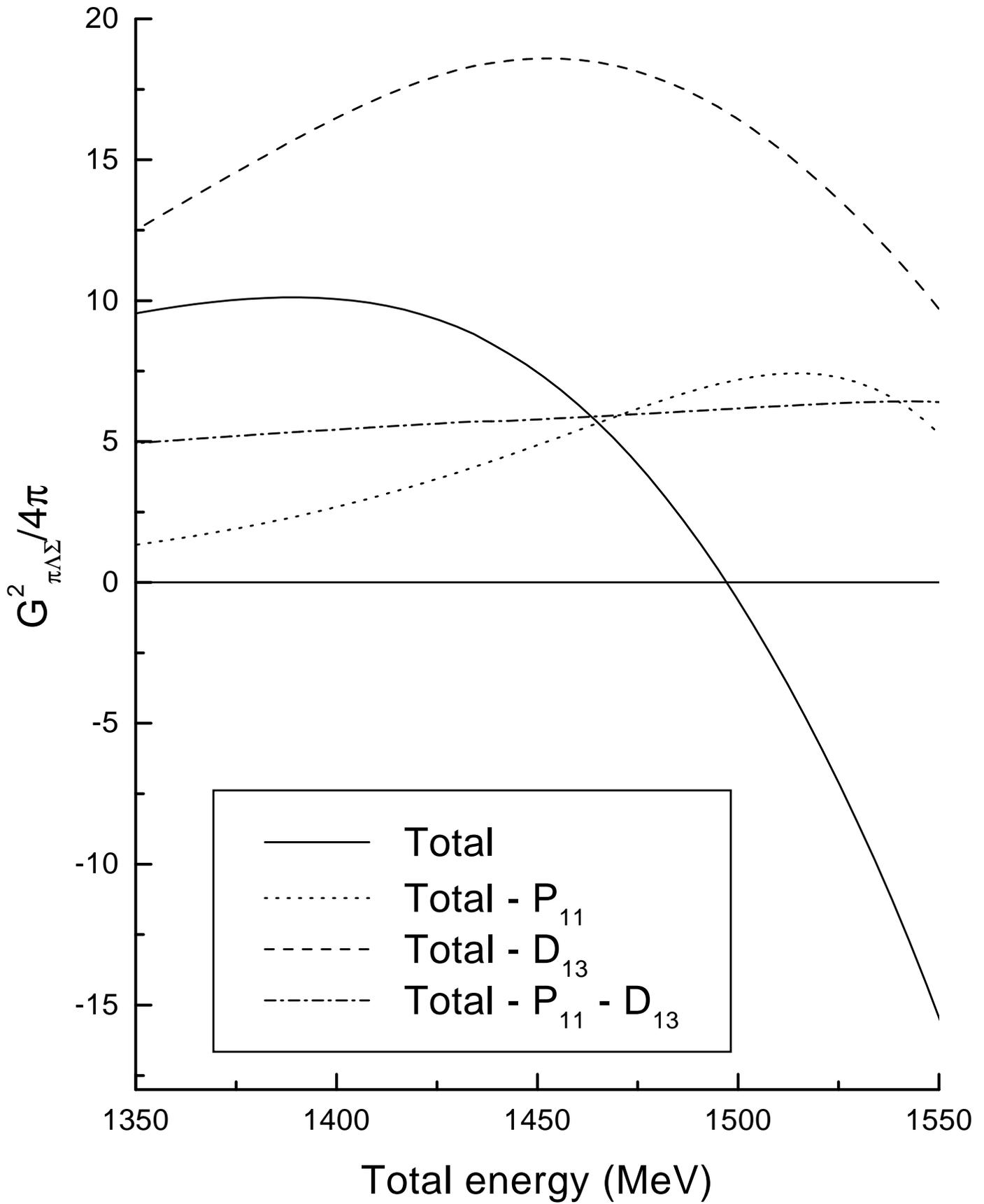
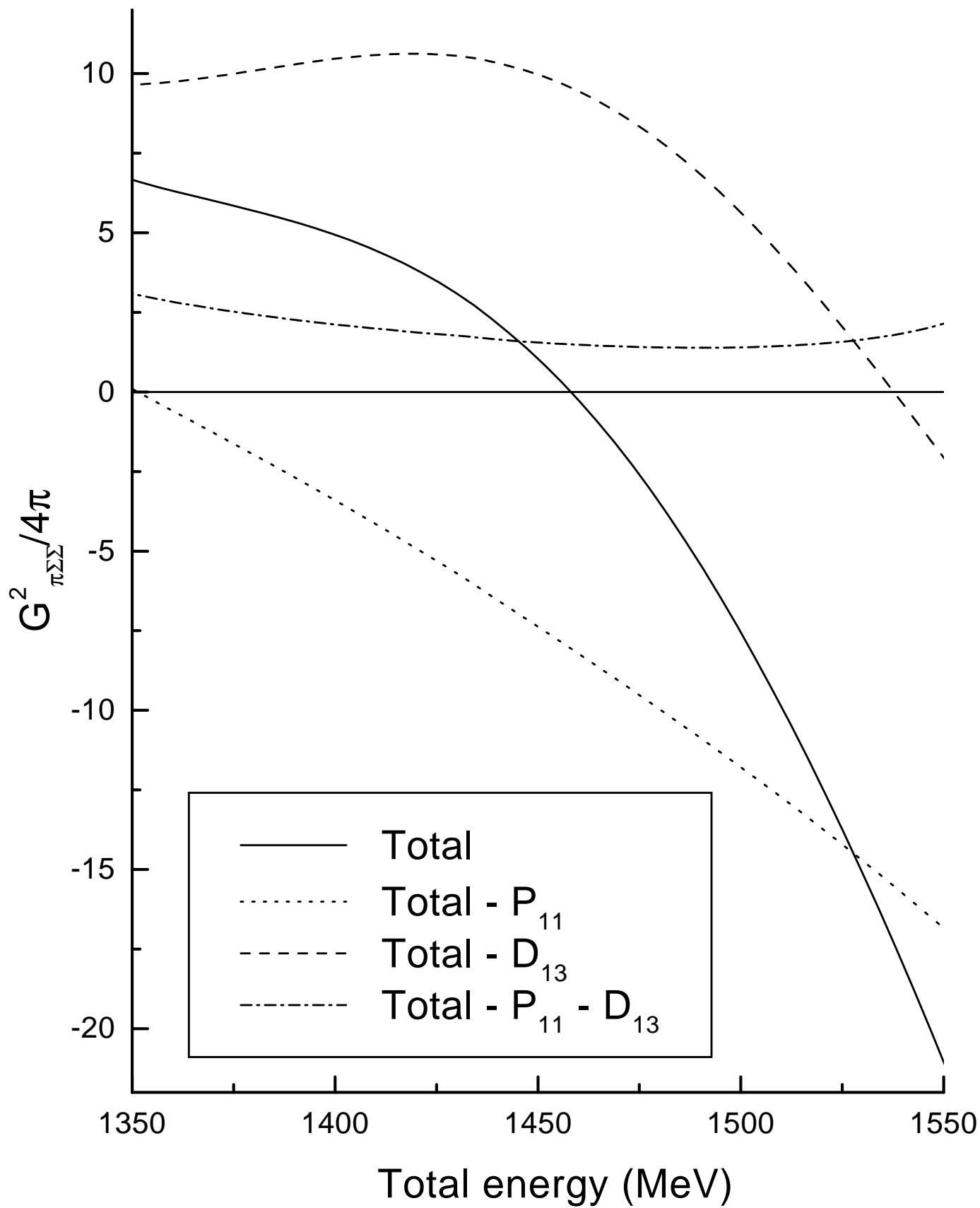
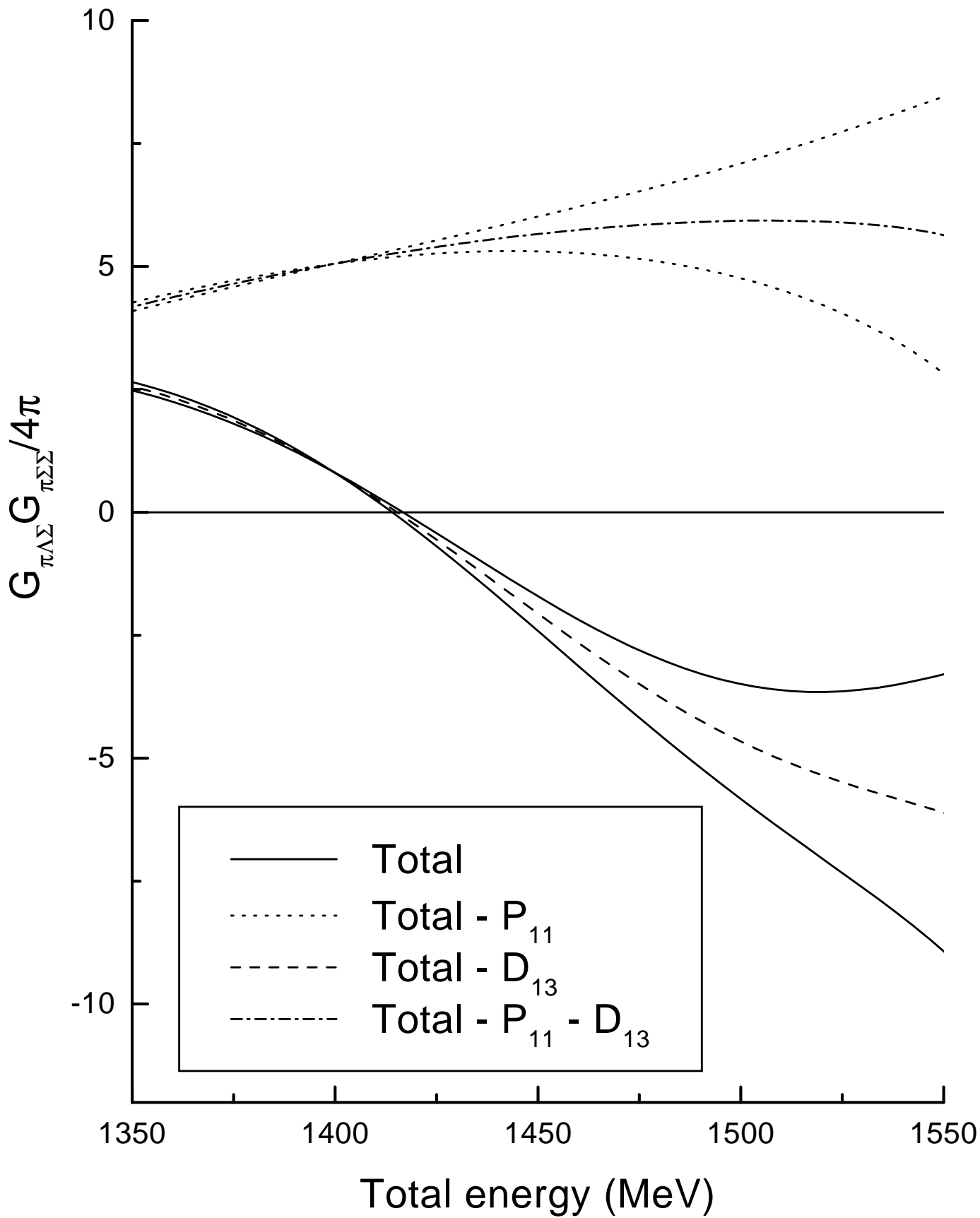
\bye